\documentclass{elsarticle}
\usepackage{graphicx}
\usepackage{amssymb}

\newcommand{\ket}[1]{\left|{#1}\right\rangle}

\newcommand{\SIGMA}{\mbox{\boldmath${\sigma}$}}
\newcommand{\ALPHA}{\mbox{\boldmath${\alpha}$}}
\newcommand{\NABLA}{\mbox{\boldmath${\nabla}$}}
\newcommand{\vc}[1]{\mathbf{#1}}

\begin{document}

\title{Simulating Dirac fermions with Abelian and non-Abelian gauge fields in optical lattices}

\author[eal]{E. Alba\corref{cor1}}
 \ead{emilio.albalinero@gmail.com}

\author[xfg]{X. Fernandez-Gonzalvo}
\author[jmp]{J. Mur-Petit}
\author[jjgr]{J.J. Garcia-Ripoll}
\author[jkp]{J.K. Pachos}

\cortext[cor1]{Corresponding author}

 \address[eal,xfg,jmp,jjgr]{Instituto de F\'\i sica Fundamental, IFF-CSIC, Calle Serrano 113b, Madrid E-28006, Spain}

\address[jkp]{School of Physics and Astronomy, University of Leeds,
  Leeds LS2 9JT, United Kingdom}


\date{\today}

\begin{abstract}
In this work we present an optical lattice setup to realize a full Dirac Hamiltonian in 2+1 dimensions. We show how all possible external potentials coupled to the Dirac field can arise from perturbations of the existing couplings of the honeycomb lattice pattern. This greatly simplifies the proposed implementations, requiring only spatial modulations of the intensity of the laser beams to induce complex non-Abelian potentials. We finally suggest several experiments to observe the properties of the quantum field theory in the setup.
\end{abstract}

\maketitle

\section{Introduction}

Lattice models have long been a powerful tool to study quantum mechanics and quantum field theories. They allow for computational treatment of analytically intractable problems, from high-Tc superconductors~\cite{TcLattice} to colour confinement in QCD~\cite{QCDLattice}. However, the computational cost of dealing with these problems can be exorbitant~\cite{Ukawa02}. In the spirit of Feynman~\cite{Feynman82}, great effort has been devoted to simulating these models in intrinsically quantum systems, which can be more efficient in reproducing key characteristics of these problems. Indeed quantum simulators have managed to produce results in the study of relativistic quantum mechanics~\cite{Gerritsma10,Gerritsma11} and quantum phase transitions~\cite{Bloch02,Friedenauer08,Kim10,Simon11}.

Optical lattices are one of the most promising candidates for quantum simulations, due to their scalability, tunability and their versatility in terms of geometry and dimensionality. They have already been successful in simulating quantum phase transitions of Bose-Hubbard and spin Hamiltonians and synthetic gauge theories~\cite{Bloch02,Simon11,Soltan11,Spielman09}, while further theoretical work has shown the possibility of addressing field theory and topological aspects of condensed matter phyisics~\cite{Jaksch03,Spielman09,Jiannis09,Mazza11,Alba11}. In this work we propose a way to study in an optical lattice the 2+1 Dirac fields coupled to both Abelian and non-Abelian gauge fields. Our work connects with previous proposals for simulating Dirac fermions in various types of lattices~\cite{Jun07,Wu08,Lee09}, including massive fermions~\cite{Lepori10} and QED simulations using BEC for the bosonic fields~\cite{Kapit11}. The novelty of this work is that instead of artificially induced phases or bosonic baths, we rely on intensity modulations of the trap and of additional laser beams to give rise to all sorts of phenomena: from the appearance of synthetic electromagnetic fields and a Dirac mass to exotic flavour-coupling perturbations. Our idea is intimately connected to the way mechanical deformations induce effective magnetic fields in graphene~\cite{Castro08,Geli10,Castro10}, but makes use of the greater tunability of optical lattices to create a larger family of potentials.

The basic ingredient in this work is a setup consisting on two state-dependent triangular lattices, connected by a Raman laser. The spatial modulations of the intensity of the laser give rise to a position-dependent hopping which, as we show, is equivalent to a deformation of the lattice. The combination of this with other lattice perturbations allows us to produce effective Abelian fields (magnetic and electric potentials), non-Abelian fields (also known as flavour coupling terms), scalar fields and an effective mass. This setup is particularly interesting because it also helps us in the detection of the gauge fields, using the tools from Ref.~\cite{Alba11}, but the ideas put forward in this work can also be generalized to other lattices with Dirac fermions, such as the honeycomb lattice or the kagome lattice, which have been recently demonstrated in the lab~\cite{Jo11}.

This work is organized in a self-contained manner, evolving from well know developments (Sections~\ref{sec:dirac-equation}-\ref{sec:lattice-distortion}) all the way up to the final implementation and detection. In section~\ref{sec:dirac-equation} we review how the Dirac equation arises from a tight-binding treatment of the honeycomb lattice. We pay special attention to the tools for deriving the continuum limit and to the appearance of an extra degree of freedom, usually called flavour~\cite{Geli10}. In section~\ref{sec:lattice-distortion} we analyse the effect of perturbations on the lattice, interpreting the result as a coupling between the Dirac field and scalar, Abelian and non-Abelian external gauge fields. In section~\ref{sec:traps} we present in detail our proposal for the experimental implementation of the Dirac model and of the external potentials using ultracold atoms in an optical lattice. After discussing a trapping scheme based on state-dependent optical lattices, section~\ref{sec:hopping-changes} shows how to implement the effective potentials from section~\ref{sec:lattice-distortion} using intensity modulations of the laser beams. Section~\ref{sec:wannier} elaborates on the experimental feasibility of the trapping scheme, estimating the Hamiltonian parameters and interaction strengths from band structure calculations. Following the setup, section~\ref{sec:measurement} discusses different experimental protocols to characterize and measure the previous quantum simulations. Finally, in section~\ref{sec:discussion} we discuss our proposal, relating it to recent proposals in the fields of optical lattices and graphene.

\section{From tight-binding models to Dirac fields}
\label{sec:dirac-equation}

Here we develop a theoretical framework in which 2D fields arise as the continuum approximation to a quadratic tight-binding model on a honeycomb lattice. Later in section~\ref{sec:traps} we show how these Hamiltonians describe non-interacting atoms in a strong periodic confining potential, but in this section the focus is on the abstract model. More precisely, we are interested in the derivation of honeycomb lattice band structure, whose excitations behave as relativistic Dirac particles, and how this procedure is modified to obtain a coupling between the Dirac particles and emergent gauge fields.

\subsection{Tight-binding model for two coupled sub-lattices}

\begin{figure}
  \centering
  \includegraphics[width=0.65\textwidth]{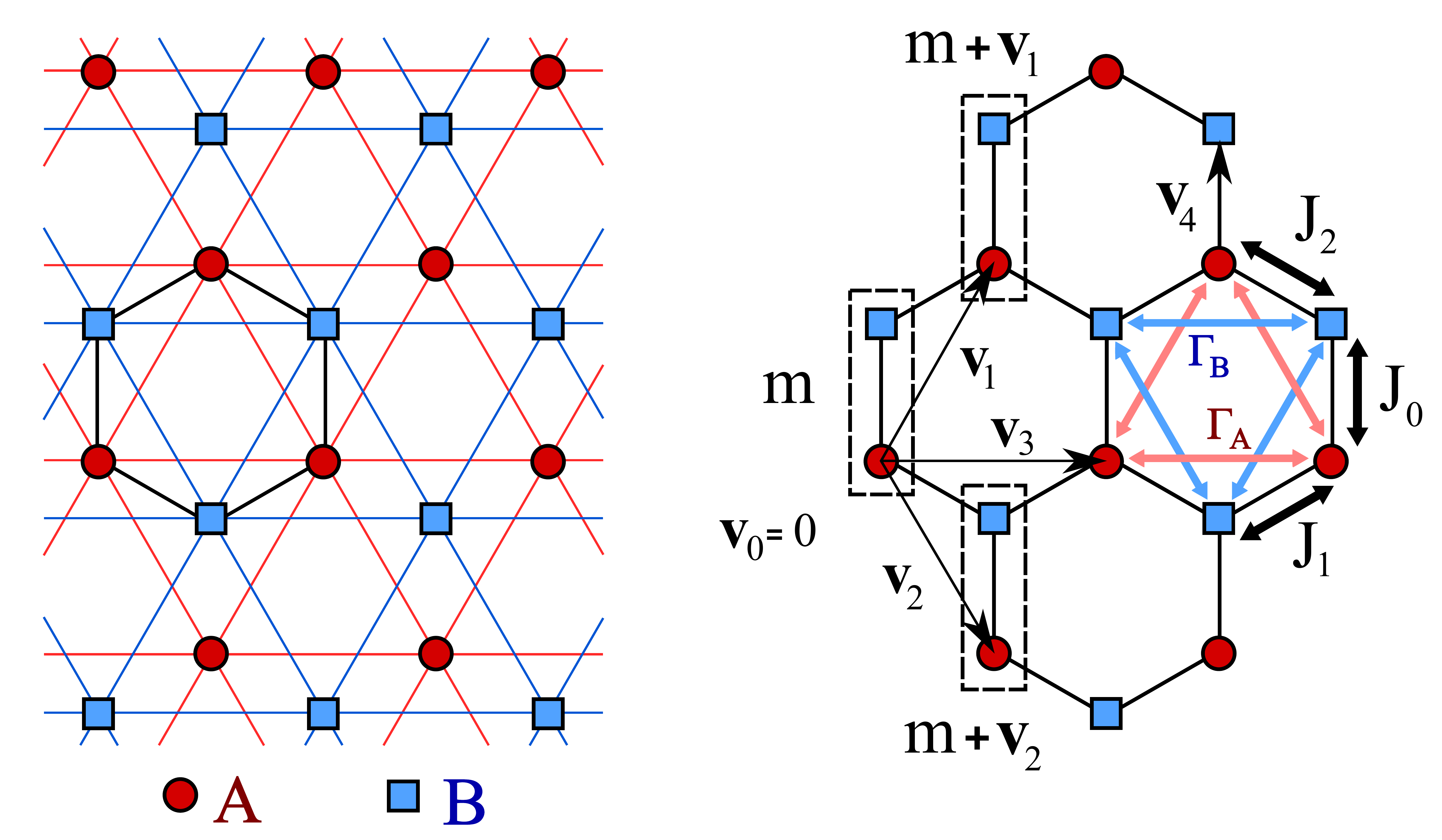}
 \caption{Two triangular sub-lattices displaced one with respect to the other to generate a honeycomb lattice. The vectors $\vc{v}_1$ and $\vc{v}_2$ are the hexagonal lattice generator vectors, whose length is the triangular lattice period, $d=|\vc{v}_{2,3}|$. In addition we define $\vc{v}_0 =  0$ and $\vc{v}_3 = \vc{v}_1 + \vc{v}_2$. The unit cells are shown with dashed rectangles, and are labelled by $m$. $\Gamma_A$ and $\Gamma_B$ are the hopping parameters related with jumps between positions inside sub-lattices $A$ and $B$ while $J_i$ are the hopping parameters related with hexagonal jumps between sublattices, each one related with the corresponding $\vc{v}_i$, $i=0,1,2$.}
  \label{Lattices}
\end{figure}

Consider a system of fermionic particles in a strong confining periodic potential. We model these particles with fermionic creation and annihilation operators which describe the presence or absence of a fermion in a particular lattice site. The lattice is bipartite, which means that it can be divided into two disjoint sets of sites $A$ and $B$, where the nearest neighbours of any $A$-site are all elements of $B$, and viceversa [cf. figure~\ref{Lattices}a)]. We consider nearest-neighbour hoppings between different sublattices of a bipartite lattice, and also next-to-nearest-neighbour (or intra-lattice) hoppings. For the hexagonal spatial geometry shown in figure \ref{Lattices}, our model Hamiltonian reads:
\begin{eqnarray}
  H &=& \sum_{m} \left( J_{0} a^{\dag}_{m}b_{m} +
  J_{1} a^{\dag}_{m}b_{m-v_{1}} + J_{2} a^{\dag}_{m}b_{m+v_{2}} 
  +\mathrm{H.c.} \right) \label{ham}\\
  &+& \sum_{v=v_1}^{v_3} \left( \Gamma_A a^{\dag}_{m}a_{m+v} +
    \Gamma_B b^{\dag}_{m}b_{m+v} +\mathrm{H.c.} \right), \nonumber
\end{eqnarray}
where the $J_i$ are the hopping parameters related with the nearest neighbour hoppings; $\Gamma_A$ and $\Gamma_B$ model the tunneling effect inside sublattices $A$ and $B$ respectively; $a^{\dag}_m$ and $a_m$ ($b^{\dag}_m$ and $b_m$) are the creation and annihilation operators for an atom in the position $m$ of sublattice $A$ ($B$), and the $v$'s refer to the $\vc{v}$ vectors represented in figure \ref{Lattices}b. The first line of equation~(\ref{ham}) represents the hexagonal hoppings, and the second line represents the two kinds of triangular hoppings. Note also that $J_0,J_1,J_2, \Gamma_A$ and $\Gamma_B$ have units of energy in this notation.

\begin{figure}[t]
  \centering
  \includegraphics[width=0.8\textwidth]{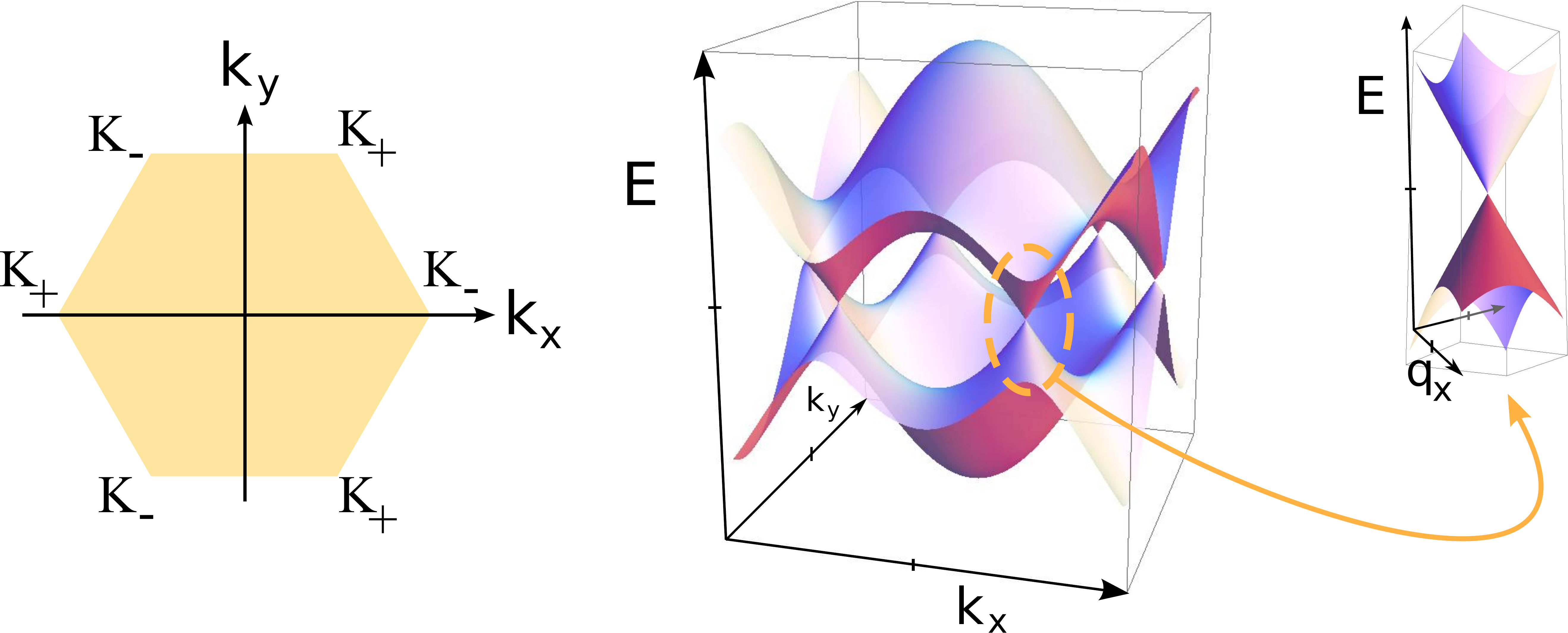}
  \caption{\emph{Left}: First Brillouin Zone (1BZ) for an hexagonal lattice. The $K_\pm$ points written in black are the non-equivalent Dirac Points. \emph{Center}: three-dimensional representation of the energy bands of the honeycomb lattice. The two surfaces represent the positive and the negative energy solutions equation~(\ref{ener}) meeting together in the Dirac points where $E=0$. \emph{Right}: zoom around the Dirac points where the energy bands take conical shape and the Hamiltonian of the system can be approximated as a Dirac Hamiltonian, being $E\propto |\vc{q}|$, equation~(\ref{dlham}).}
  \label{EnergyThings}
\end{figure}

\subsection{Energy bands and Dirac field}

The honeycomb Hamiltonian is recovered from equation~(\ref{ham}) by making $J_0=J_1=J_2 \equiv J$ and $\Gamma_A=\Gamma_B=0$: \begin{equation}
  H_{hon}= J \sum_{m} \left\{ a^{\dag}_{m}b_{m} +a^{\dag}_{m}b_{m-v_{1}} +a^{\dag}_{m}b_{m+v_{2}} + H.c. \right\}.
  \label{gham}
\end{equation}
To calculate the energy bands, we do a Fourier transform over the creation and annihilation operators as
\begin{equation}
  a^{\dag}_{m} = \int\limits_{\Omega} \frac{\mathrm{d}^2k}{2\pi} e^{i \vc{k} \cdot \vc{r}_{m}} a^{\dag}_{\vc{k}},
  \label{ft}
\end{equation}
where $\vc{k}$ is the momentum, $\vc{r}_m$ is the position of the site $m$, the integration limit $\Omega$ makes reference to the first Brillouin Zone (1BZ) and a similar expression is used for $b$. The diagonalization of the Hamiltonian in momentum space leads to the eigenvalues of the energy that form the energy band structure; namely
\begin{equation}
  E_\pm(\vc{k})= \pm J \sqrt{ 3 + 4 \cos{\left(\frac {1}{2}k_{x} d\right)} \cos{\left(\frac{\sqrt{3}}{2} k_{y} d \right)} + 2\cos{\left( k_{x} d \right)}},
  \label{ener}
\end{equation}
which are represented in figure \ref{EnergyThings}. There are six points in the 1BZ for which $E=0$, but only two of them are non-equivalent and we name them $K_{\pm}=(\mp\frac{4\pi}{3},0)/d$, where $d$ is the {\em triangular} lattice spacing~\cite{Semenoff84}. Assuming only small energy perturbations around $E=0$, the excitations will be confined to the so called Dirac cones. To find the effective theory for these low-energy and long-wavelength excitations one makes an expansion around the Dirac points of $E(\vc{k})$, introducing $\vc{q} = \vc{k} - \vc{K}_\pm$ and assuming that for the relevant states it remains ``small'' (i.e. $|\vc{q}| \times d \ll 1$). Under these conditions, it is possible to work with the Dirac cones as if they extended over all values of $\vc{q}$, with integrals in $\mathrm{d^2 k}$ over the 1BZ being replaced by an integral in $\mathrm{d^2q}$ over the whole momentum space. By transforming the operators back to position space via inverse Fourier transform
\begin{equation}
  \phi_a^\dagger(\vc{r}) = \int \frac{1}{2\pi} a^{\dag}_{\vc{K} + \vc{q}} e^{i (\vc{K} + \vc{q}) \cdot \vc{r}} \mathrm{d^{2} q},
  \label{ift}
\end{equation}
one gets an effective theory for the continuum fields $\{\phi_a(\vc{r}), \phi_b(\vc{r})\}$, satisfying the Dirac-like Hamiltonian (cf. equation \ref{dham})
\begin{equation}
  h_+(\vc{q}) \propto \SIGMA \cdot \vc{q},
  \quad h_-(\vc{q}) = -h_+(\vc{q})^\star,
  \label{dlham}
\end{equation}
where $\SIGMA = (\sigma_x, \sigma_y)$ are the usual Pauli matrices and the subscript $\pm$ stands for the choice of cone ($K_\pm$).

\subsection{Derivation of the Dirac Hamiltonian in position space}

While the previous discussion reveals the main ingredients of the relativistic fields which arise from our discrete Hamiltonian, two concerns appear. One is the added difficulty of treating spatially dependent perturbations, caused by the two consecutive Fourier transforms. The other concern is the little attention paid to the fact that there are two intrinsically different cones, which in line with previous literature we call the two ``flavours'' of Dirac particles in the lattice. As we will show later, this flavour plays a role in the simulation of non-Abelian fields.

In order to extend the derivation of the Dirac dispersion relation to setups in which translational invariance is weakly broken, we rely on a continuum-field approach that bypasses the use of momentum space~\cite{Jackiw07}. Starting with our tight-binding Hamiltonian~(\ref{gham}), we approximate the Fock operators at each lattice site as the value of a continuous field defined over all space, but which varies so smoothly that it is approximately constant over each unit cell. Moreover, since we are interested in excitations around the $K_\pm$ quasimomenta, these fields are the envelope of quasi-plane wavepackets around those points
\begin{equation}
  \hat{a}_m \rightarrow \sqrt{\mathrm{d}^2r} \left( \hat{\Psi}_{a+}(\vc{r}_m) e^{-i
      \vc{K}_+ \cdot \vc{r}_m}+\hat{\Psi}_{a-}(\vc{r}_m) e^{-i
      \vc{K}_- \cdot \vc{r}_m} \right),
\end{equation}
with an equivalent expression for $\hat{b}_m$. Here $\mathrm{d}^2r$ is the unit cell area and the exponential containing the Dirac points $\vc{K}_{\pm}$ implements the desired wavepacket ansatz. Substituting these operators in the tight-binding model produces the following limit Hamiltonian
\begin{eqnarray}
  \label{Flavour Interaction Hamiltonian}
  &\hat{H} \propto \displaystyle\sum_{\tau,\zeta,i} \int e^{i\vc{r}(\vc{K}_\tau-\vc{K}_\zeta)} \left[\hat{\Psi}^{\dagger}_{a\tau}(\vc{r})
  \hat{\Psi}_{b\zeta}(\vc{r}+\vc{v}_i) e^{- i \vc{K}_\tau\cdot\vc{v}_i} + \mathrm{H.c.}\right]\,\mathrm{d}^2r,
\end{eqnarray}
where the Greek indices stand for the two possible flavour choices ($\pm$ cones). Note that the index $i=0,1,2$ runs over the three vectors $\vc{v}_0, -\vc{v}_1,\vc{v}_2$ depicted in figure~\ref{Lattices} and which connect different unit cells. This is so, because we are free to define the smoothly varying fields $\Psi_a$ and $\Psi_b$ on the middle point of the unit cell.

Since we have enforced the fields to be slowly varying, the presence of the rapidly oscillating exponential $e^{i\vc{r}(\vc{K}_\tau-\vc{K}_\zeta)}$ imposes the condition $\tau=\zeta$ and we end up with two flavour-decoupled integrals. A more quantitative argument would compare the $\tau=\zeta$ term with one that oscillates with $\vc{K}_+ - \vc{K}_-$. Since both are proportional to $J$, we only need to compare their ratio. Integrating by parts the oscillating term and using periodic or open boundary conditions we may write the rapidly oscillating term correction as
\begin{equation}
  \sum_{\tau\neq\zeta} \int e^{i\vc{r}(\vc{K}_\tau-\vc{K}_\zeta)} \left[\hat{\Psi}^{\dagger}_{a\tau}(\vc{r})
    \left(\frac{\vc{K}_\tau - \vc{K}_\zeta}{|\vc{K}_\tau-\vc{K}_\zeta|^2}
    \cdot \nabla\right)
  \hat{\Psi}_{b\zeta}(\vc{r}+\vc{v}_i) + \mathrm{H.c.}\right]\,\mathrm{d}^2r,
\end{equation}
which becomes small if the fields oscillate slowly, that is $ d |\nabla \Psi| \ll |\Psi|$.

At this point one may expand $\hat{\Psi}(\vc{r}+\vc{v}) \simeq [1+\vc{v}\cdot\NABLA]\hat{\Psi}(\vc{r})$ so the Hamiltonian reads:
\begin{equation} 
  \hat{H} \propto \sum_{\tau,i} \int \left\{ \hat{\Psi}^{\dagger}_{a \tau}(\vc{r})
  \left[ 1+\vc{v}_i \NABLA \right]\hat{\Psi}_{b \tau}(\vc{r}) e^{-i \vc{K}_\tau \cdot \vc{v}_i} + \mathrm{H.c.}\right\}\mathrm{d}^2r.
\end{equation}
Our resulting Hamiltonian is therefore diagonal both in position and in flavour space but couples the internal degrees of freedom $a$-$b$ via an off-diagonal operator with only one nontrivial term:
\begin{equation} 
  C_{\tau}=1+e^{-i\vc{K}_\tau \cdot \vc{v}_2} [1+\vc{v}_2 \cdot \NABLA]+
  e^{i \vc{K}_\tau \cdot  \vc{v}_1}[1-\vc{v}_1 \cdot \NABLA]
\end{equation}
Since $1+e^{-i\vc{K}_\tau \cdot \vc{v}_2}+e^{i \vc{K}_\tau \cdot \vc{v}_1}$ vanishes due to the definition of the Dirac cones, the coupling term simplifies to
\begin{equation} 
  C_\pm=e^{-i\vc{K}_\pm \vc{v}_2} \vc{v}_2 \cdot \NABLA-e^{i \vc{K}_\pm
    \cdot \vc{v}_1}\vc{v}_1\NABLA
  = \frac{\sqrt{3}d}{2}[\partial_y - \pm i\partial_x].
\end{equation}
Introducing a momentum operator $\vc{q}=-i\NABLA$, our coupling term becomes $C_\tau \sim \tau q_x + iq_y$ and the Hamiltonian spatial density is therefore:
\begin{eqnarray} 
  h_+(\vc{r}) = c \left( \begin{array}{cc} 0 & C_+ \\ C^{\star}_+ &
      0 \end{array}\right) &=c \left( \begin{array}{cc} 0 & q_x + iq_y
      \\ q_x - iq_y & 0 \end{array}\right)= c\,\vc{q} \cdot \SIGMA
  \nonumber \\ h_-(\vc{r}) = -h_+^*(\vc{r}) 
\end{eqnarray}
which is the expression of the massless Dirac Hamiltonian with Fermi velocity $c=\sqrt{3}Jd/2\hbar$. Note that changing flavour is equivalent to changing the sign of $q_x$ in the Hamiltonian.

\section{Extending the Dirac Hamiltonian by modifying the lattice parameters}
\label{sec:lattice-distortion}

We now generalize the free Dirac Hamiltonian to include a variety of external fields~\cite{DiracBook}
\begin{equation}
  H_D=c \, \ALPHA \cdot \vc{p} + \beta mc^2 + \beta V_{cov}.
  \label{gdham}
\end{equation}
Here $m$ is the mass of the particles and $V_{cov}$ is the most general covariant potential containing scalar, vector, matrix, pseudoscalar, pseudovector and pseudotensor fields. The Dirac matrices $\alpha^i=\gamma^0\gamma^i$ and $\beta=\gamma^0$, are defined in terms of the generators of the Clifford group, $\gamma^{\mu}.$ In 2+1 dimensions this is a set of $2\times 2$ matrices satisfying the anticommutation relations $\{\gamma^{\mu}, \gamma^{\nu}\}=2 \eta^{\mu\nu}$ with $\eta^{\mu\nu}=diag(1,-1,-1)$, where $\mu,\nu=0,1,2$. The choice $\ALPHA=\SIGMA=(\sigma_x,\sigma_y)$ and $\beta=\sigma_z$ corresponds directly to the Hamiltonian in equation~(\ref{dlham}). The Dirac Hamiltonian then has the form
\begin{equation}
  H_D= c\,\SIGMA \cdot \left( \vc{p} - e \vc{A}(\vc{r}) \right) + e A^0(\vc{r}) \, \mathbb{I} + \left( mc^2 + V(\vc{r}) \right)\! \sigma_z,
  \label{dham}
\end{equation}
where $A^0(\vc{r})$ and $\vc{A}(\vc{r})$ are the usual scalar and vector potential that give rise to observable electric and magnetic fields, and $V(\vc{r})$ is a scalar potential that mimics the effect of an imposed mass; $e$ is an effective charge. Note that since the fields $\vc{A}$ and $A^0$ do not have any dynamics, we may assume $e=1$, for convenience.  Let us now show how to recover all terms in equation~(\ref{dham}) by slightly perturbing the tight-binding model.

\subsection{Generating a mass term} \label{massderivation}

The simplest term that we can add to Hamiltonian equation~(\ref{gham}) is an energy difference between atoms in sublattices $A$ and $B$:
\begin{equation}
  \delta H =\sum_{m} \left( \frac{\varepsilon}{2}  a^{\dag}_{{m}} a_{{m}} - \frac{\varepsilon}{2}  b^{\dag}_{{m}} b_{{m}} \right), 
\end{equation}
which in the continuum limit becomes
\begin{equation}
  \delta H =  \sum_\tau \int \left(\frac{\varepsilon}{2}\hat{\Psi}^{\dagger}_{a \tau}(\vc{r}) \hat{\Psi}_{a \tau} (\vc{r})-\frac{\varepsilon}{2}\hat{\Psi}^{\dagger}_{b \tau}(\vc{r}) \hat{\Psi}_{b \tau} (\vc{r})\right)\mathrm{d}^2r
\end{equation}
or equivalently
\begin{equation}
  \delta h (\vc{r}) \sim \sigma_{z} \frac{\varepsilon}{2}.
\end{equation}
Comparing this with equation~(\ref{dham}), we see that the position-independent term proportional to $\sigma_z$ can be identified with an effective mass.

\subsection{Abelian potentials}

We can extend our model by starting from equation~(\ref{gham}) and smoothly modifying the hoppings as $J_{i} =J + \epsilon_{i,m} $ where $|\epsilon| \ll | J|$:
\begin{equation}
  \delta H =\sum_{m,i} \epsilon_{i,m}  a^{\dag}_{{m}} b_{{m+v_i}} + \mathrm{H.c.} 
\end{equation}
In the continuum limit this renders
\begin{equation}
  \delta H =\sum_{\tau,i} \int e^{i \vc{K}\cdot \vc{v}_i}\epsilon_i(\vc{r})\hat{\Psi}^{\dagger}_{a \tau} (\vc{r}) \hat{\Psi}_{b \tau} (\vc{r}+\vc{v}_i) \mathrm{d}^2r.
\end{equation}
In the notation of the previous section, this is equivalent to a change in the ``coupling term'' between pseudospins:
\begin{equation}
  \delta C_\tau = \frac{\sqrt{3}}{2} \left(\tau \frac{\epsilon_0}{J}-\tau\frac{\epsilon_1+\epsilon_2}{2J} + i \frac{(\epsilon_2-\epsilon_1)}{J}\right)
\end{equation}
therefore allowing us to formally derive an Abelian external potential
\begin{equation}
  H =c\, \SIGMA \cdot \left( \vc{q}-\vc{A}(\vc{r}) \right)
\end{equation}
where $\vc{A} =(\tau \mathrm{Re} [\delta C_\tau], \mathrm{Im} [\delta C_\tau])$.

\subsection{Scalar fields}

Consider now perturbations which are diagonal in the internal space to the tight-binding Hamiltonian, such as those given by intralattice hoppings in equation~(\ref{ham}):
\begin{equation}
  \delta H =\sum_m \sum_{i=1}^3 \Gamma_A  a^{\dag}_{{m}} a_{{m+v_i}} + \Gamma_B  b^{\dag}_{{m}} b_{{m+v_i}}+h.c. 
\end{equation}
Following the continuum limit performed in section \ref{massderivation} the Hamiltonian in momentum space around the $K_\tau$ point can then be written as
\begin{equation}
  H'_\tau= c\,\SIGMA\cdot\vc{q} - \left(
    \begin{array}{cc}
      \sum_i e^{-i \vc{K}_\tau\cdot\vc{v}_i}\Gamma_A(\vc{r}) & 0 \\
      0 & \sum_i e^{-i \vc{K}_\tau\cdot\vc{v}_i}\Gamma_B(\vc{r}) 
    \end{array}
  \right),
  \label{hprima}
\end{equation}
or purposefully rewritten (reabsorbing the constant $\sum_i e^{-i \vc{K}_\tau \vc{v}_i}$ flavour-dependent factor) as
\begin{equation}
  H'_\tau= c\,\SIGMA\cdot  \vc{q}  - \frac{(\Gamma_A(\vc{r})+\Gamma_B(\vc{r}))}{2} \mathbb{I} - \frac{(\Gamma_A(\vc{r})-\Gamma_B(\vc{r}))}{2} \sigma_z
\end{equation}
which has both a term proportional to $ \mathbb{I}$ representing an electric potential, $\phi _\pm (\vc{r}) = \sum_i e^{-i \vc{K}_\pm \vc{v}_i}\frac{1}{2}[\Gamma_A(\vc{r})+\Gamma_B(\vc{r})]$, and a term proportional to $\sigma_z$ which contributes to the effective mass.

\subsection{Flavour-coupling perturbations}

We have seen that spatial variations of the hopping elements emerge as an Abelian external field. These variations have to be \textit{small}, but it should be stated that, in order to keep flavours decoupled, they also have to be \textit{slowly varying}. Otherwise, if there is a modulation with wavevector comparable to the order $(\vc{K}_+-\vc{K}_-)$, the flavour-coupling terms $\tau \neq \zeta$ in equation~(\ref{Flavour Interaction Hamiltonian}) do not vanish. What we suggest now is to introduce small perturbations whose wavelength is comparable to the lattice constant and thus bridge the difference in momentum between cones. For instance
\begin{equation}
  \label{Simple perturbation}
  \delta \hat{H}= \sum_{m,i} 2 \chi_{x,i,m}
  \cos \left[ \vc{r}\cdot(\vc{K}_+-\vc{K}_-)\right] a^\dagger_m b_{m+v_i}
\end{equation}
with a constant coupling strength $ \chi_{x,i,m}=\chi_x$. These rapid oscillations only allow the survival of terms that couple different cones, cancelling all terms inside the same cone. The most general perturbation of this sort is 
\begin{equation}
  \delta\hat{h}(\vc{r})=\chi_x(\vc{r}) \left(\hat{\Psi}^{\dagger}_{a+}\hat{\Psi}_{b-}
    +\hat{\Psi}^{\dagger}_{a-}\hat{\Psi}_{b+} + \mathrm{H.c.} \right),
\end{equation}
where $\chi_x$ is the spatial dependence of the slow envelope that surrounds our perturbation~(\ref{Simple perturbation}). Actually, this envelope can be ``remodulated'' in order to make flavour coupling also weakly spatial-dependent. Moreover, we can also introduce in equation~(\ref{Simple perturbation}) sine terms, $\epsilon(\vc{r})=\chi_y \sin \left[\vc{r}\cdot(\vc{K}_+-\vc{K}_-)\right]$, which make the flavour-coupling term complex.

\subsection{The complete Hamiltonian}

Since there exist no pseudo-potentials ($\gamma^5=1$ in $2+1$ dimensions), our idea of perturbing the tight-binding model parameters allows us to reconstruct all possible external potentials of the Dirac equation, plus an additional coupling between different types of particles. With all these elements we have the following effective single-particle Hamiltonian:
\begin{equation}
  h(\vc{r}) =\left( \begin{array}{cc} c\,\SIGMA\cdot[\vc{q}-\vc{A}(\vc{r})]+m\sigma_zc^2 +\phi_+(\vc{r})  &
      \chi_x \sigma_x + \chi_y \sigma_y\\ \chi_x \sigma_x + \chi_y \sigma_y  &
      -c \{\SIGMA\cdot [\vc{q}-\vc{A}(\vc{r})]\}^*+mc^2\sigma_z +\phi_-(\vc{r}) \end{array}\right) .
\end{equation}

\section{Trapping of atoms in an optical honeycomb lattice}
\label{sec:traps}

In this section we introduce optical lattices and explain how to construct a honeycomb lattice with two state-dependent triangular optical lattices. We discuss how to implement a tight-binding Hamiltonian and propose an experimental setup to obtain the Dirac field and the external fields using solely Raman and detuned lasers.

\subsection{From two triangular sub-lattices to a honeycomb lattice}

Recent advances in the development of high-aperture objectives and their integration in optical traps open the door to the generation of almost arbitrary two-dimensional potential landscapes for ultracold atoms. The basic idea is that off-resonant light may be used to tightly confine atoms in the maxima or minima of intensity, recreating sophisticated lattice models~\cite{Jaksch98}. While until now those minima and maxima were generated through the interference of multiple laser beams~\cite{Bloch02,Jaksch98}, a novel paradigm consists on shaping and organizing those intensity profiles by simply projecting sophisticated images on the two-dimensional focal plane of a lens. The first experiments along this line have reproduced the usual square lattice quantum simulations~\cite{Greiner09} and also demonstrated the first triangular lattices~\cite{Chin10}.

\begin{figure}
  \centering
  \includegraphics[width=0.3\textwidth]{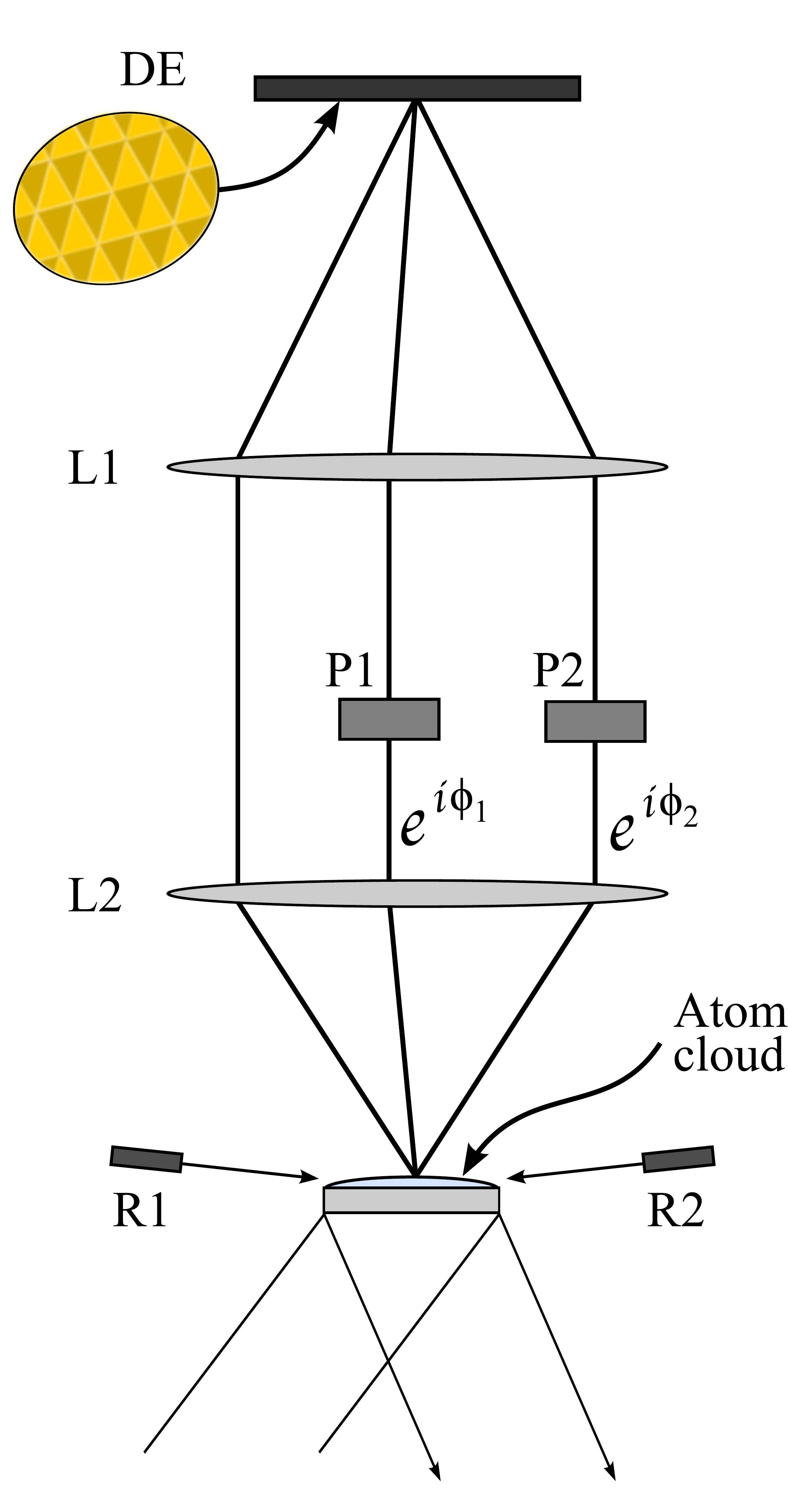}
  \caption{Proposal of experimental setup. DE is the diffraction element, L1 and L2 are the required lenses to make the beams parallel and to focalize them on the lattice plane, R1 and R2 are the lasers used for the Raman transitions, $P_1$ and $P_2$ are phase plates used to change the relative phases, $e^{i\phi_{1,2}}$, of the $\sigma^+$ and $\sigma^-$ beams.}
  \label{Setup}
\end{figure}

In this work we are particularly interested in the last of those setups, which combines two triangular lattices~\cite{Chin10} in the same plane. In this experiment the trapping laser beams are first diffracted by a holographic mask with a triangular pattern, selecting the  first diffraction orders, which are then collected by a powerful lens to create the imprinted intensity pattern at its focal plane. The relative phases of the three diffracted beams may be independently controlled in a way that allows the displacement of the resulting triangular lattice (cf. figure~\ref{Setup}). When this procedure is applied to two laser beams that differ in frequency or polarization, it becomes possible to produce two triangular lattices, $A$ and $B$, which coexist on the same plane and have a tunable relative separation. The result is the original setup introduced in figure~\ref{Lattices} in an abstract way, where now $A$ and $B$ are physically implemented by an optical potential.

In order to jump from two independent triangular lattices to the setup introduced in figure~\ref{Lattices}, we need means to introduce the couplings between different lattice sites, that is hoppings from one lattice to another, or within the same lattice. The most direct way to implement this in an optical system is to use the two lattices $A$ and $B$ to trap atoms of the same species but in different internal states, $\ket a$ and $\ket b$ respectively. We also need a way to rotate between $\ket a$ and $\ket b$, which can be implemented via Raman transitions. In a situation with all these ingredients two kinds of hoppings are allowed for the atoms: on the one hand an atom in sub-lattice $A$ ($B$) can tunnel between positions in its own sub-lattice, as contemplated in our model by the hopping parameter $\Gamma_A$ ($\Gamma_B$). This parameter can be controlled by increasing or decreasing the intensity $I_A$ ($I_B$) of the laser beam generating each sublattice. On the other hand a Raman-induced change in the internal state of an atom from $\ket a$ to $\ket b$ (from $\ket b$ to $\ket a$) will make the atom shift from sublattice $A$ to sublattice $B$ ($B$ to $A$). These nearest-neighbour jumps can be different in each of the three possible directions, and are modelled by the hopping parameters $J_0$, $J_1$ and $J_2$ which are associated with the vectors $\vc{v}_0$, $\vc{v}_1$ and $\vc{v}_2$ shown in figure~\ref{Lattices}.

\subsection{Realization of two state-dependent triangular sub-lattices}

How do we implement in practice the ideas from the previous subsection, and in particular the coupling between lattices? The engineering and control of state-dependent lattices is a mature technology~\cite{Jaksch98,Mandel03}, which nevertheless requires some careful control of the atomic states and decoherence. We will briefly describe how this works for fermionic alkaline atoms and how this integrates with the projected lattices scheme.

\begin{figure}
  \centering
  \includegraphics[width=0.5\textwidth]{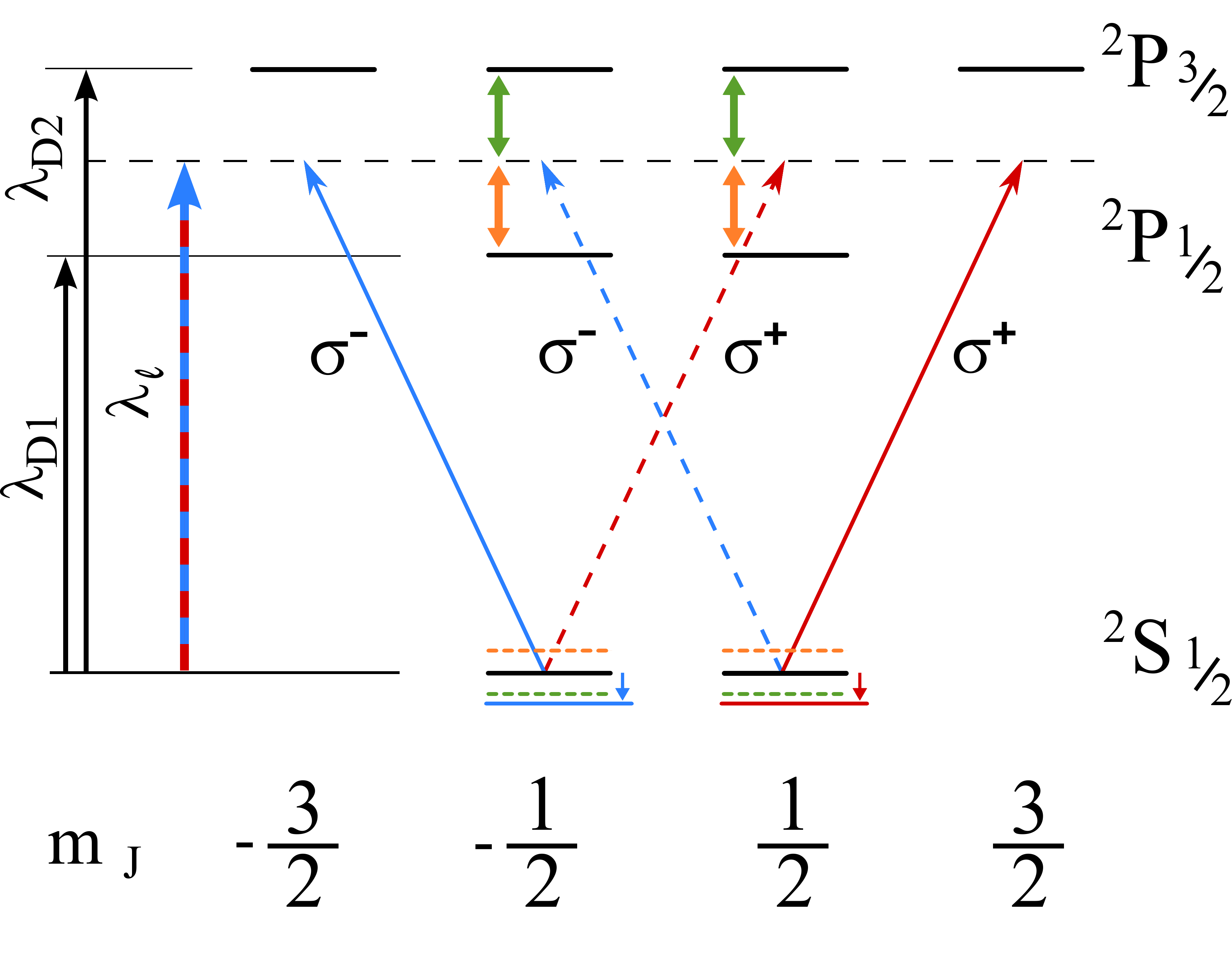}
  \caption{Fine-structure energy levels of an alkaline atom. \emph{Not to scale}. $\lambda_{D1}$ and $\lambda_{D2}$ are the wavelengths of the D1 and D2 lines respectively. $\lambda_l$ is the wavelength of the optical lattice lasers. Dashed lines mark the cancelled ac-Stark effects: a $\sigma^+\,(\sigma^-)$ polarised photon incident on a $\ket{L=0, J=\frac{1}{2}, m_J=-\frac{1}{2}}\;\left(\ket{0, \frac{1}{2}, +\frac{1}{2}}\right)$ state would see both the $\ket{1,\frac{1}{2},\frac{1}{2}}$ and the $\ket{1,\frac{3}{2},\frac{3}{2}}\;\left(\ket{1,\frac{1}{2},-\frac{1}{2}} \;\mathrm{and}~\ket{1,\frac{3}{2},-\frac{3}{2}}\right)$ states in such a way that the ground level energy displacements cancel with each other. The net effect is then only due to $\sigma^-\,(\sigma^+)$, marked with continuous lines.}
  \label{EnrgLvls}
\end{figure}

As sketched in figure~\ref{EnrgLvls}, it is possible to find a wavelength $\bar \lambda_l$ falling between the $D1$ and $D2$ lines of an alkaline atom for which polarised light only traps atoms in one of the ground states manifolds. More precisely, the fine-structure energy levels of alkaline atoms are denoted by $\ket{L, J, m_J}$, where $L$ is the electron angular momentum quantum number, $\vec J$ the combined orbital and spin momentum and $m_J$ is its projection along the quantization axis. When we illuminate with $\sigma^+$ polarised light at a frequency $\omega=\frac{1}{2}(\omega_{D1}+\omega_{D2}),$ the ac-Stark shift that it induces on the $\ket{0,\frac{1}{2},-\frac{1}{2}}$ cancels due to the positive and negative contribution of off-resonant $D1$ and $D2$ transitions. The result is that circularly polarized $\sigma^\pm$  light can only trap atoms in the $\ket{0,\frac{1}{2},\pm\frac{1}{2}}$ states, respectively.

\begin{figure}
  \centering
  \includegraphics[width=0.6\textwidth]{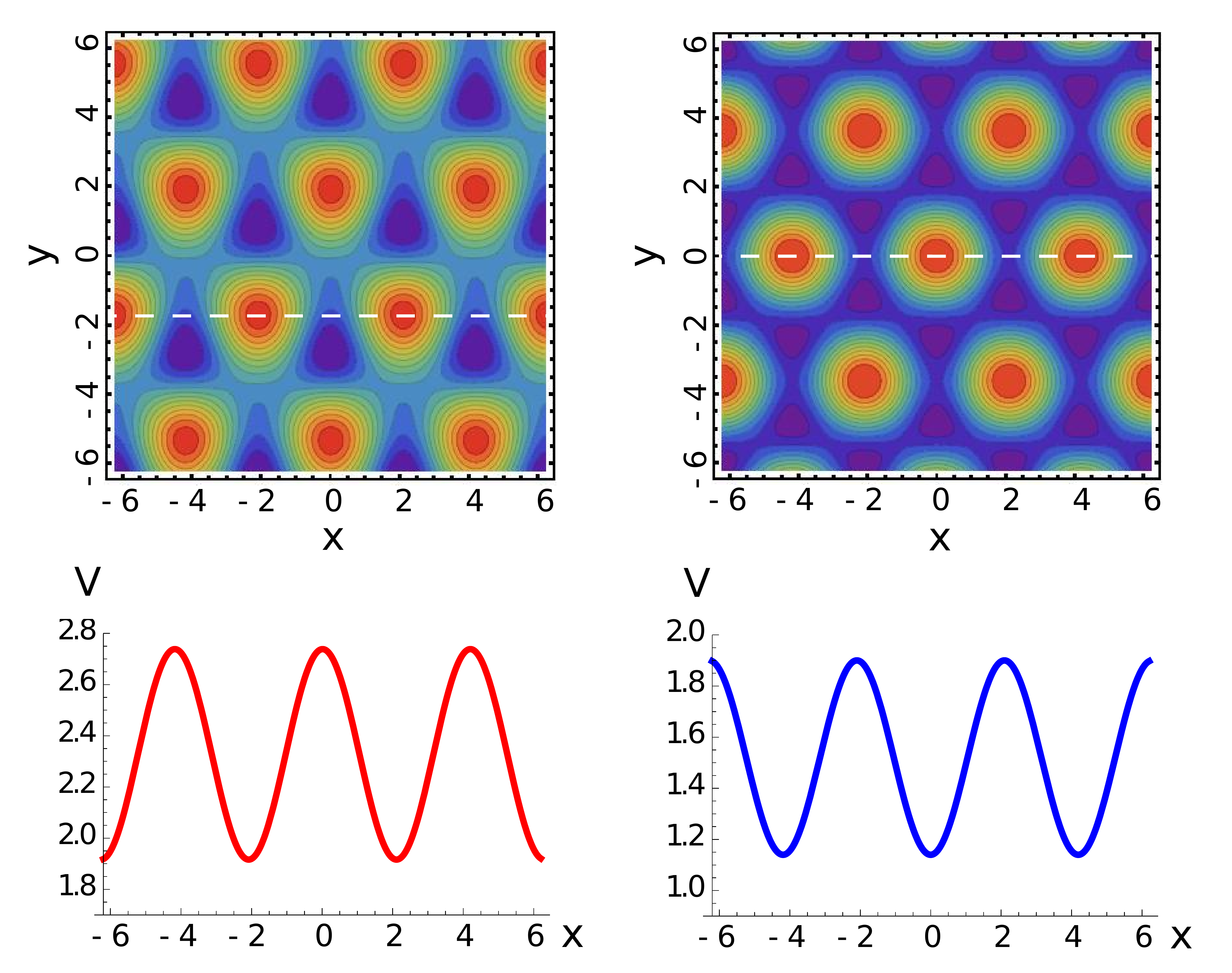}
 \caption{\emph{Left}: sub-lattice generated by $V_A=\frac{2}{3}V_+ + \frac{1}{3}V_-$. \emph{Right}: sub-lattice generated by $V_B=V_-$. \emph{Top}: trapping potentials as functions of $x$ and $y$, in arbitrary units. The red zones represent the potential minima where the atoms are trapped. It is apparent that both sub-lattices are triangular and displaced one with respect to the other. By superimposing both images one obtains the structure of an hexagonal lattice. \emph{Bottom}: transversal cut corresponding to the white dashed line along the $x$ direction. The difference in potential depth can be controlled by changing the ratio $V_+/V_-$, but this will affect the relative positions of the potential minima.}
  \label{LatFuncs}
\end{figure}

In practice, however, the situation is more subtle because atoms also have some hyperfine structure, induced by the coupling between the electronic and nuclear angular momenta. Let us focus on the fermionic species $^6$Li, in line with previous proposals~\cite{Chin10}. Out of the hyperfine ground-states $\ket{F,m_F}_{HF}=c_1\ket{J=\frac{1}{2},m_J=\frac{1}{2}}+c_2\ket{J=\frac{1}{2},-\frac{1}{2}}$ where $\vec F = \vec I + \vec J$ being $\vec I$ the nuclear angular momentum of $^6$Li ($I=1$), we select $\ket{a}=\ket{\frac{1}{2},-\frac{1}{2}}_{HF}=-\sqrt{\frac{2}{3}}\ket{\frac{1}{2},\frac{1}{2}}+\frac{1}{\sqrt 3}\ket{\frac{1}{2},-\frac{1}{2}}$ and $\ket{b}=\ket{\frac{3}{2},-\frac{3}{2}}_{HF}=1 \ket{\frac{1}{2},-\frac{1}{2}}$. If $V_+$ and a $V_-$ are the intensity distributions that result by illuminating with light in the $\sigma^+$ and $\sigma^-$ polarisations, the states $\ket a$ and $\ket b$ will feel the ac-Stark potentials $V_A=\frac{2}{3} V_+ + \frac{1}{3} V_-$ and $V_B= V_-$. As shown in figure~\ref{LatFuncs}, the combined potentials can look like two displaced triangular lattices by choosing the adequate relative phases between the diffracted beams. Furthermore, the relative depths can be controlled by changing the ratio $V_+/V_-$, but this may require a new tuning of the relative phases.

The last ingredient is a coupling between both lattices. This can be done using a Raman laser that couples both internal states, $\ket a$ and $\ket b$. When an atom in state $\ket a$ is affected by the laser, it will switch state, but since the energy must be conserved, this implies tunneling to a neighboring site in the other lattice. This qualitative description assumes that atoms in the $A$ and $B$ sublattices are confined to the lowest energy band. A very important question is whether this tight-binding approximation is compatible with the Raman laser. This will be discussed in the following section.

\subsection{Band structure calculations}
\label{sec:wannier}

Our experimental proposal relies on the possibility to combine two different lattices and couple them via Raman assisted tunneling. The feasibility of this procedure has been experimentally demonstrated in a superlattice experiment~\cite{Aidelsburger11}. It has also been discussed at length in the various works that suggest implementing gauge fields via photon-assisted tunneling, beginning with the seminal paper by Jaksch and Zoller~\cite{Jaksch03}, and also in a related work on coupled 1D lattices~\cite{Garciaripoll07}. Despite this, it is very illustrative to do a quantitative discussion of the lattice parameters involved in this setup, with the aim of clarifying what hopping, $\Gamma_{A,B}$ and coupling strengths, $J_i$, can be achieved, and what is the limit of weak perturbations that we will rely on later in the manuscript.

Our basic tool in this discussion is the expansion of the field operator in terms of Wannier wavefunctions~\cite{Jaksch98}. We will assume that the two triangular lattices are defined using a similar potential, $V_{\mathrm{triang}}(\vc{x})$, which for the sake of concreteness we choose
\begin{equation}
  V_{\mathrm{triang}}(\vc{x}) = V_0 \left[3 - 2\cos(2\pi x) \cos(2\pi y/\sqrt{3})
    - \cos(4\pi y/\sqrt{3})\right].
\end{equation}
This potential gives rise to two single-particle Hamiltonians, with a relative displacement given by $\vc{v}_4$(see figure~\ref{Lattices}):
\begin{eqnarray}
  H_{A} &=& -\frac{\hbar^2}{2m}\NABLA^2 + V_{\mathrm{triang}}(\vc{x}/d),\\
  H_{B} &=& -\frac{\hbar^2}{2m}\NABLA^2 + V_{\mathrm{triang}}[(\vc{x}-\vc{v}_4)/d],
\end{eqnarray}
whose eigenstates are the Bloch waves $\psi_{\vc{k}}(\vc{x})$. The Wannier functions are sums over these Bloch states in a Brillouin zone, $ w(\vc{x}) = \frac{1}{|\mathcal{B}|^{1/2}} \int \psi_{\vc{k}}(\vc{x}) \mathrm{d}^2k$ and we assume that they have the same shape for both lattices. Furthermore, we assume that there is a strong confining harmonic potential in the z-direction such that the atoms are restricted to its vibrational ground state with an approximate length scale $(\hbar/m\omega)^{1/2}\sim d$.

\begin{figure}
  \centering
  \includegraphics[width=0.7\linewidth]{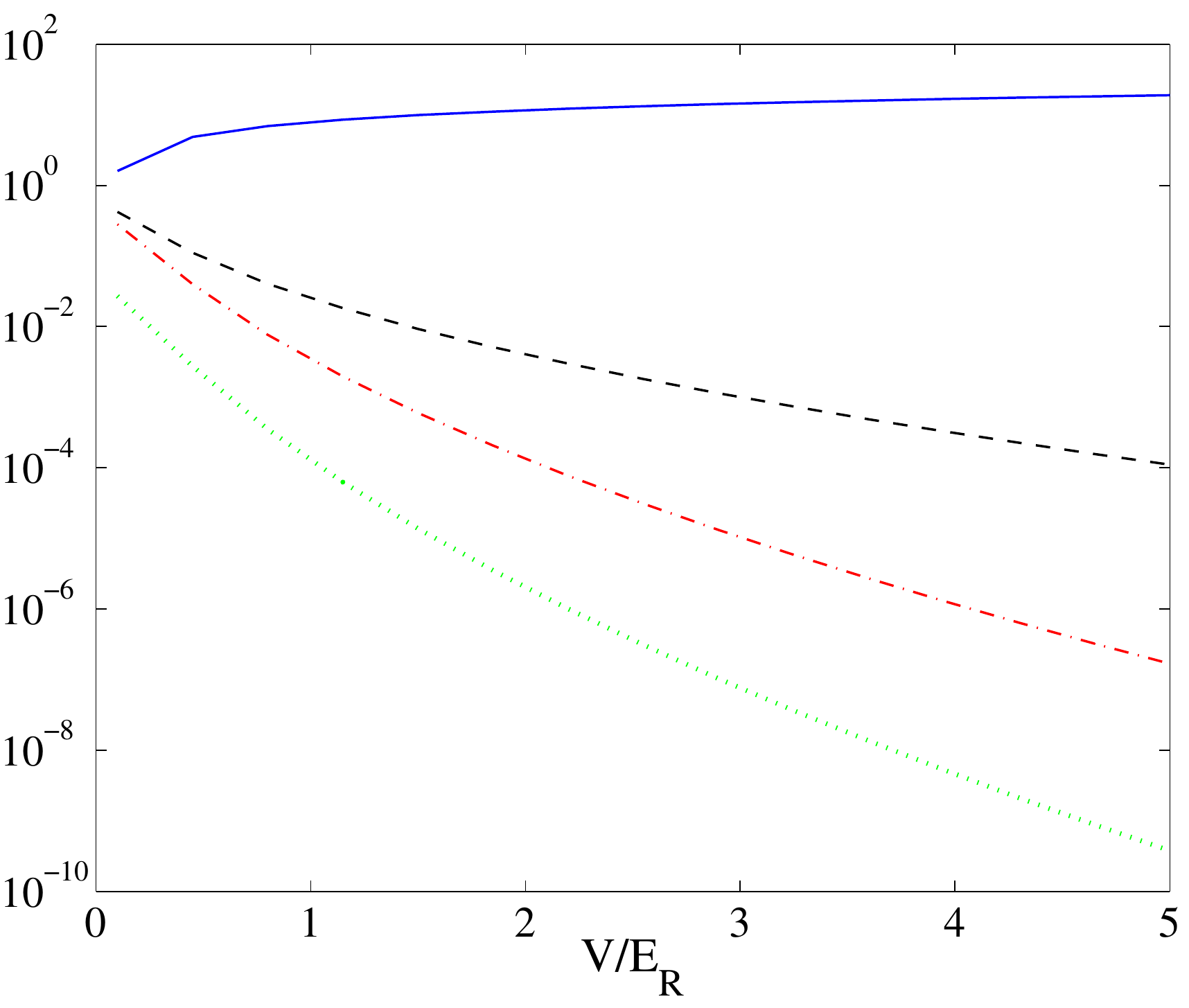}
  \caption{Dependence of the tight-binding parameters on the intensity of the confining lattice potential. We have numerically calculated the values of the on-site interaction strength $Ud^3/g$ (solid, blue), the hexagonal tunneling amplitude $J/\Omega$ (black, dashed), the interlattice interaction stregth $U_{AB}d^3/g$ (red, dotted-dashed) and the triangular tunneling amplitude $\Gamma{A,B}/E_R$ (green, dotted).}
  \label{fig:wannier}
\end{figure}

In the low energy limit that is usual for these experiments, the fermionic field may be approximated by a linear combination of these localized states
\begin{equation}
\label{TBBands}
  \hat{\psi}_a(\vc{x})^\dagger \simeq \sum_m \hat{a}_m^\dagger w(\vc{x}-\vc{x}_m),\;
  \hat{\psi}_b(\vc{x})^\dagger \simeq \sum_m \hat{b}_m^\dagger w(\vc{x}-\vc{x}_m-\vc{v}_4).
\end{equation}
In this limit, the motion of particles is described by a tight-binding Hamiltonian, where the  intralattice hopping amplitude is defined as the nearest-neighbor term of the Hamiltonian using this expansion, or
\begin{equation}
  \Gamma_A = \int w(\vc{x}-\vc{v}_{1,2,3})^\star H_{A} w(\vc{x}) \mathrm{d}^2x,
\end{equation}
and similarly for $\Gamma_B$. If we also include a Raman coupling among lattices, characterized by a Rabi frequency $\Omega$, we may still use the tight-binding approximation provided that this frequency is much smaller than the gap between the lowest energy band and the first excited band of the triangular lattice, $|\Omega| \ll \Delta E$. If this is the case, we will have that the interlattice coupling can be estimated as
\begin{equation}
  J = \Omega \int w(\vc{x}-\vc{v}_4)^\star w(\vc{x}) \mathrm{d}^2x.
\end{equation}
Note that for equation~(\ref{TBBands}) to remain valid, $\Omega$ has to be small when compared with $V_0$, not with the actual hoppings. This means that the ratio between $J$ and $\Gamma_{A,B}$ is not fixed, and that one hopping does not have to be small with respect to the other one.

We have computed numerically the Wannier functions for a triangular lattice setup using a discretization of the Brillouin zone with 100 modes and expanding the Bloch wave with up to 625 modes. Using this we have estimated the integrals corresponding to the hoppings, and also to the on-site and nearest-neighbor interactions, given by
\begin{eqnarray}
  U &=& \frac{g}{d}\int_{\mathcal{C}} |w(\vc{x})|^4 \mathrm{d}^2x\\
  U_{AB} &=& \frac{g}{d}\int_{\mathcal{C}} |w(\vc{x})|^2|w(\vc{x}-\vc{v}_4)|^2 \mathrm{d}^2x,
\end{eqnarray}
where $\mathcal{C}$ denotes the surface of the unit cell. The results are shown in Fig.~\ref{fig:wannier}, where we used the recoil energy $E_R = 4\pi^2\hbar^2/2md^2$ as unit of energy to ease the comparison, typical values for the lattice spacing ($400-600$ nm), and interaction strength values for alkaline fermions which range from $g/d^3\simeq0.01 E_R$ ($^{40}K$) to $g/d^3\simeq0.1 E_R$ ($^{6}Li$)~\cite{Julienne10}. Note how the ratio $J/\Omega$ is at least 100 times larger than $\Gamma_{A,B}/E_R$ for reasonable values of the potential depth. This means that even if we have to impose $\Omega \ll V_0$, we still can reach a regime in which $J \simeq \Gamma_{A,B}$. Note also that while the interlattice term $U_{AB}$ decreases rapidly, there is still a window of values where it might be comparable to the influence of hopping, opening the door to experiments with gauge fields \textit{and} interactions.

\section{Perturbing the hopping parameters}
\label{sec:hopping-changes}

In the previous discussion we presented a feasible setup to obtain a honeycomb tight-binding Hamiltonian with neutral atoms in an optical lattice. In this section we take the scheme one step further so that the hopping parameters can be locally changed to incorporate mass terms and pseudofields to the Dirac Hamiltonian. We propose to optically control the Dirac field physics much in the way mechanical strains have been suggested in graphene sheets~\cite{Castro08,Geli10,Castro10}. Two proposals to implement variations are presented, the first of which is best in line with the setup shown in the previous section, the other being a more formal approach to general hopping calculations across potential wells.

\subsection{Spatial dependence of Raman intensity}
\label{sec:raman-distortion}

Interlattice hoppings are assisted by an external laser beam inducing Raman transitions between the internal states of the atoms, thus creating a transition amplitude which scales with the intensity of the beam. By changing the spatial dependence of this amplitude one can overprint a first-neighbours hopping perturbation $\epsilon_i(\vc{r})$. This method is the easiest to implement in this setup, but is very limited and unable to act on intralattice (next-to-nearest-neighbour) transitions. Therefore it is the chosen method to add Abelian vector potentials and non-Abelian coupling terms.

\subsection{Lattice distortions}
\label{sec:distortion}

A more general analysis can be made by considering spatial distortions of the lattice, that is, either through a relative displacement of one site with respect to its neighbor or by adjusting the width of the individual potential wells. The displacement may be achieved by controlling the relative phases of the conforming beams through the phase modulators \textit{P1} and \textit{P2} in figure~\ref{Setup}. The width of the potential wells can be modified by a change in the height of the confining potential, using a mask which modulates the intensity of the laser beams. We now develop a simple model for calculating the influence of these two modifications in the hopping parameters.

We assume that the ideal (unperturbed) hopping parameters between wells behave as if lattice sites were spatially separated harmonic oscillators in an original reference frame $\vc{r}'$. The lattice distortion described locally by the transformation $\vc{r}'=A\vc{r}$ renders the following single-well Hamiltonian:
\begin{equation}
  \hat{H}=\frac{- \hbar^2}{2 m} \Delta + \frac{m \tilde{\omega}^2}{2} \vc{r}^TA^TA\vc{r}
\end{equation}
where the frequency of the site trap is also changed $\omega \rightarrow \tilde{\omega}(\vc{r})$, due to modulations of the intensity of the confining beam (the kinetic energy terms are not transformed since they are expressed in ``real'' space, while the potential term is meant to look like a perturbed harmonic oscillator). The localized Wannier function in the well is
\begin{equation}
  \psi_{GS}(x)=\frac{1}{\sqrt{N}} \exp (-\frac{1}{2 \sigma^2} \vc{r}^TB\vc{r})
\end{equation}
where $B=\sqrt{A^TA}$, $\sigma^2=\hbar/m \tilde{\omega}$ and $N=\pi \sigma^2/\sqrt{|B|}$. Notice that $B$ is positive-definite by construction and $\sigma$ is $\vc{r}$-dependent. If we assume no vibrational levels can be excited, the hopping parameter between first-neighbour wells separated by a vector $\vc{a}$ is simply proportional to the overlap between wavefunctions
\begin{equation}
  J \sim \int \psi^*_{GS}(\vc{r}) \psi_{GS}(\vc{r}-\vc{a})  \mathrm{d}^2r=
  \exp(-\frac{\vc{a}^TB\vc{a}}{4 \sigma^2}).
\end{equation}

This will be valid as long as variations are smooth and small, corresponding to the wavefunctions being good approximations to the ground state of the potential well and the restriction to the lowest vibrational level. Therefore the dislocation must be small: $A \simeq \mathbb{I}+ \epsilon C \rightarrow G=A^TA \simeq \mathbb{I} + \epsilon (C+C^T)$, defining $C$ as our generator. Let us call the unperturbed vector connecting both wells $\vc{a}_0$, so that the actual vector becomes $\vc{a}=A^{-1}\vc{a}_0=(\mathbb{I}-\epsilon C)\vc{a}_0$. The bilinear matrix expands as $B=\sqrt{G} \simeq \sqrt{\mathbb{I} + \epsilon (C+C^T)} \simeq \mathbb{I} +\frac{1}{2} \epsilon (C+C^T)$. Substituting these values yields:
\begin{eqnarray}
  J &\sim &\exp(-\frac{\vc{a}^TB\vc{a}}{4 \sigma^2}) \Rightarrow J=J_0 \exp ({-\epsilon \delta}),
  \nonumber \\
  \delta &=& -\frac{m \tilde{\omega}(\vc{r})}{8 \hbar} \vc{a}_0^T (C+C^T))\vc{a}_0 .
\end{eqnarray}
This result can be readily interpreted. There are two contributions: one ($C+C^T$) which modifies the hopping due to the change in distance between wells, and another ($\tilde{\omega}(\vc{r})$) which refers to the intensity profile of the confining laser beam. We can therefore tune these parameters, e.g. displacing or rotating one sublattice on top of the other, to allow for variations of the tight-binding hoppings. As explained before, all these changes are used to simulate Abelian and non-Abelian external gauge fields.

\section{Experimental detection}
\label{sec:measurement}

The most direct consequence of the appearance of Abelian and non-Abelian external fields is the distortion of the energy bands, ranging from the movement of the Dirac cones in momentum space to the appearance of a gap in the spectrum. Two methods are proposed here to observe these changes: the first one is the measurement of the momenta after removing the confining potential so as to explore the form of the energy bands; the second one focuses on measuring how these fields affect the dynamics of a single particle (or small group of particles) moving in the lattice.

\subsection{Time of flight images}

It is possible to probe the atomic population within the first Brillouin zone (BZ) of an optical lattice. The procedure consists of the following steps: (i) adiabatically switching off the lattice potentials so that the atoms quasimomenta are converted into real momenta, (ii) then letting the atoms expand freely during a certain time of flight and (iii) finally taking an absorption image of the expanded cloud~\cite{Kohl05}.

\begin{figure}
  \centering
  \includegraphics[width=0.5\textwidth]{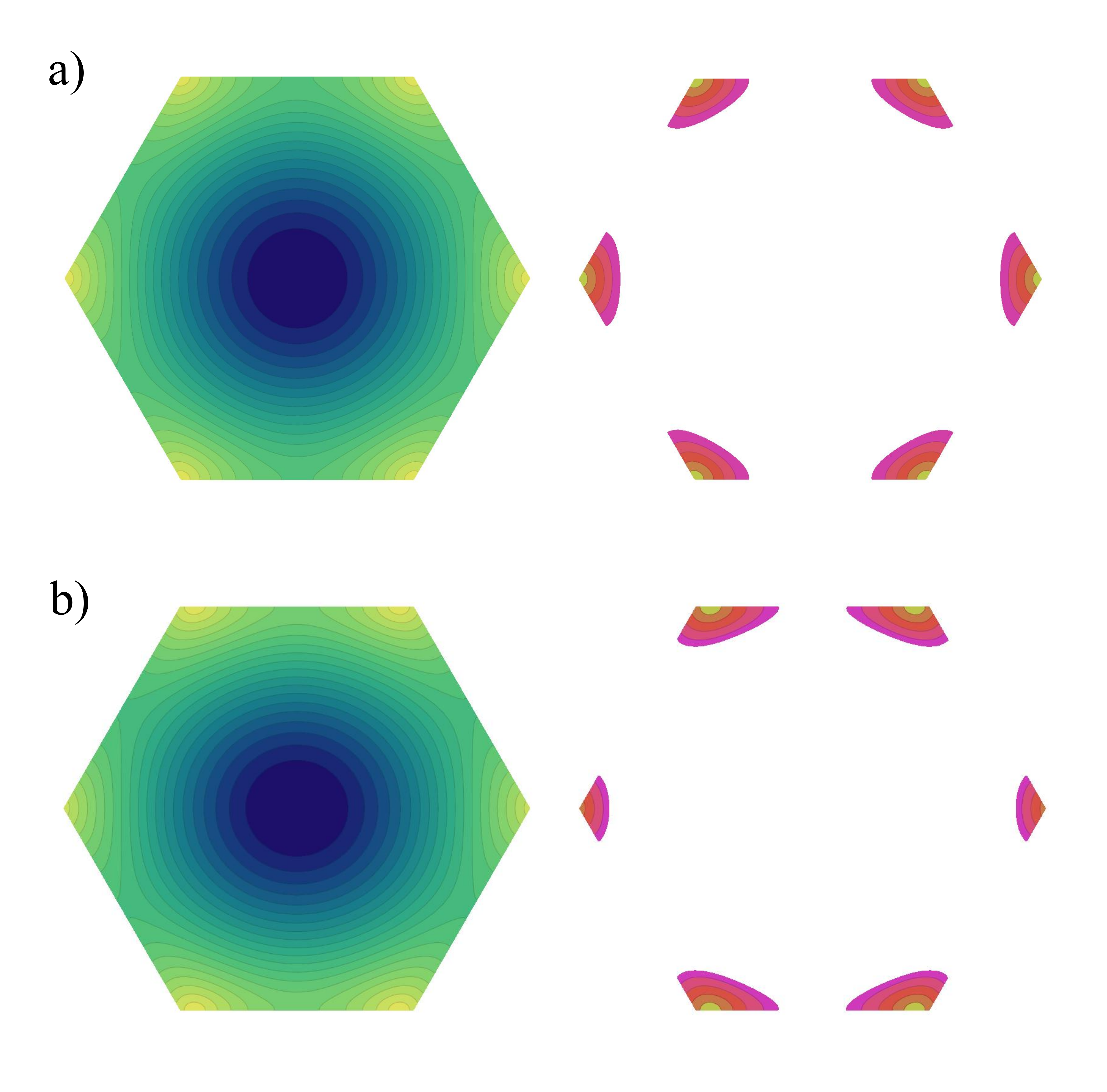}
  \caption{Expected momenta distribution for $B$ (\emph{left}) and $A$ (\emph{right}) atoms associated with the $E_-$ and $E_+$ energy bands respectively (see text). The two \emph{top} images correspond to the unperturbed honeycomb tight-binding Hamiltonian, and the \emph{bottom} ones correspond to the \emph{Abelian potential} case described in section \ref{sec:lattice-distortion} due to a perturbation in $J_0$.}
  \label{bzs}
\end{figure}

In our setup the situation is a bit more complicated. Let us denote by  $|{+,\vc{k}}\rangle$ and $|{-,\vc{k}}\rangle$ the eigenstates associated with upper and lower energy bands, $E_+(\vc{k})$ and $E_-(\vc{k})$, of the tight-binding Hamiltonian~(\ref{gham}). In a ground state with a Fermi energy slightly above zero, we would have many atoms in the lower band sector occupying the whole of the first BZ and only few atoms in the upper band, concentrated around Dirac points (See figure~\ref{bzs}a). Despite the difference, by means of absorption images one can not distinguish between $\ket +$ and $\ket -$ states as both have components corresponding to the $\ket a$,$\ket b$ internal states of the atoms.

\begin{figure}
  \centering
  \includegraphics[width=0.4\textwidth]{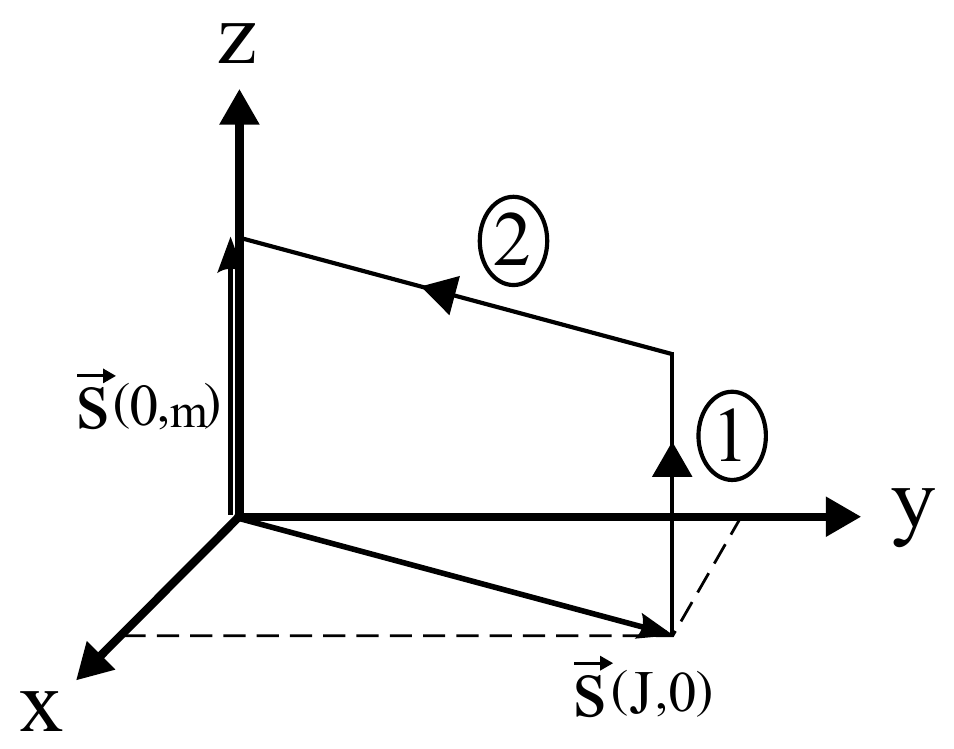}
  \caption{Adiabatic path for a transformation of the $\ket +$, $\ket -$ states into $\ket a$, $\ket b$ states. The hamiltonian of the system is written as $H \sim \vec s (J,m) \cdot \vec\sigma$, so $\vec s_{initial}=\vec s (J,0)$ lies in the $x,y$ plane while $\vec s_{final}=\vec s (0,m)$ lies along the $z$ axis. During the first step $m$ is adiabatically increased, and during the second $J$ is adiabatically decreased until zero.}
  \label{apath}
\end{figure}

In order to picture the Dirac cones we need a method that discriminates between energy bands, say, transforming all the $\ket +$ states into $\ket a$'s and all the $\ket -$ states into $\ket b$'s, while preserving the momentum, $\vc{k}$. The adiabatic theorem provides us the way for doing this. Our starting point is the honeycomb lattice Hamiltonian in momentum space
\begin{equation}
  \tilde H = J \sum_{\vc{k}} u^{\dag}_{\vc{k}}
  \left[ f(\vc{k}) \sigma^+ + f^\star (\vc{k} ) \sigma ^- \right]
  u_{\vc{k}},
  \label{mham}
\end{equation}
written with the pseudospin structure $u_{\vc{k}} = (a_{\vc{k}}, b_{\vc{k}})$, the couplings $f(\vc{k}) = 1 + e^{i \vc{k} \cdot \vc{v}_1} + e^{-i \vc{k} \cdot \vc{v}_2}$, and the Pauli ladder operators $\sigma^\pm$. Note how the Hamiltonian is a composition of commuting terms for each value of the momentum, $\vc{k}$. We will adiabatically distort all terms, following the route in figure~\ref{apath}, which consists on first adding a mass term, $\sim m\sigma^z,$ and then decreasing $J$ down to zero. The protocol maps the two eigenstates of the initial Hamiltonian, $\ket +$ and $\ket -$, to $\ket a$ and $\ket b$, accurately.

This technique would allow for the experimental observation of the effects of an Abelian potential described in section~\ref{sec:lattice-distortion}, as represented in figure \ref{bzs}b, where the Dirac cone displacement manifests as a deformation of the $\ket a$-momentum distribution. In general all other effects, which are summarized in Table~\ref{tab}, could be observed by similar methods.

\begin{table}
\caption{summary table of all the considered perturbations of the hopping parameters with their physical effects.}
\label{tab}
\small
 \begin{tabular}{@{}|p{4cm}|p{5.9cm}|p{2cm}|}
\hline
Lattice modification&Hamiltonian modification&Effect \\
\hline
\hline
\hbox{$J_{i}=J+\epsilon_i$ \ \ \ \ \ \ \ i=0,1,2}
\hbox{small hopping parameter} modification&
\hbox{$H \rightarrow c \SIGMA \cdot (\vc{q} - \vc{A})$}
$\vc{A}=(\tau \mathrm{Re} [\delta C_\tau], \mathrm{Im} [\delta C_\tau])$&
Abelian field \\
\hline

$+\sum_{m}\left( \frac{\varepsilon}{2}a^{\dag}_{m}a_{m}-\frac{\varepsilon}{2}b^{\dag}_{m}b_{m}\right)$&{$H\rightarrow c\SIGMA \cdot \vc{q}+\frac{\varepsilon}{2}\sigma_{z} $}&{Mass term}\\
\hbox{\strut energy difference between} $A$ and $B$&&\\
\hline

$\Gamma_{A} , \Gamma_{B}$
\hbox{intrasublattice triangular} hoppings&
$H\rightarrow c\SIGMA\cdot\vc{q} + \frac{1}{2} (\Gamma_{A}+\Gamma_{B}) \mathbb{I} +\frac{1}{2} (\Gamma_{A}-\Gamma_{B})\sigma_{z}$&
\hbox{Scalar field}\\
\hline

$\epsilon (\vc{r}) =\chi_x \cos ( \vc{r} (\vc{K}_+-\vc{K}_-))$&$\delta\hat{h}(\vc{r})=\chi_x(\hat{\Psi}^{\dagger}_{a+}\hat{\Psi}_{b-}
  +\hat{\Psi}^{\dagger}_{a-}\hat{\Psi}_{b+} + \mathrm{H.c.})$& \hbox{Flavour} coupling \\
\hline
  \end{tabular}
\end{table}

\subsection{Spin textures in time-of-flight images}
\label{sec:textures}

\begin{figure}
  \centering
  \includegraphics[width=\textwidth]{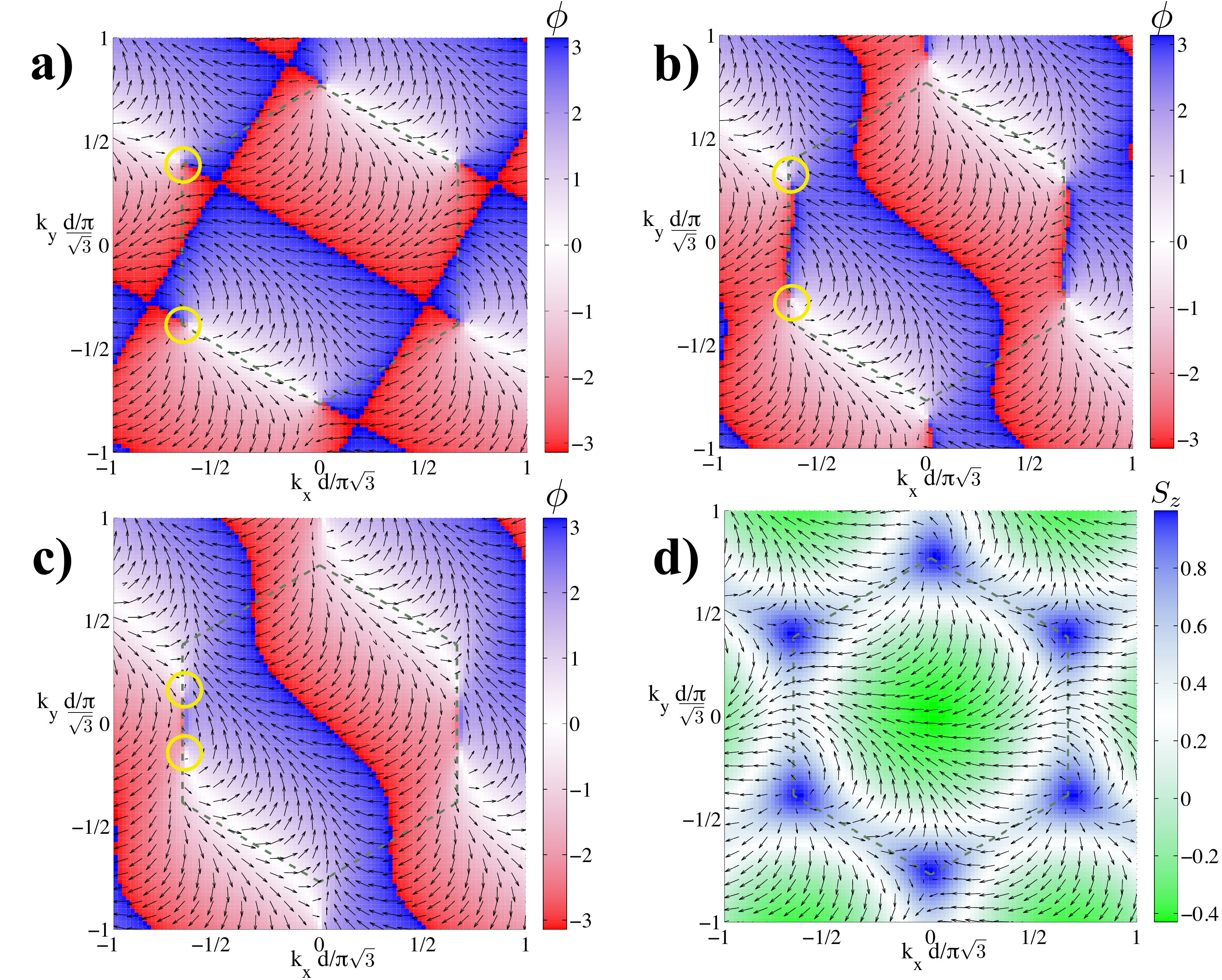}
  \caption{a),b),c): Effect of the Abelian potential by the spin texture method. In color scale, $\phi=\tan^{-1} (S_y/S_x)$. The yellow circles show the position of the Dirac cones for $\vc{A}=\vc{0},(1/\sqrt{48},0)/d,(1/\sqrt{12},0)/d$ respectively. d) Effect of a non-zero intralattice hopping ($\Gamma_A \neq 0$). In all panels, the $(S_x,S_y)$ vector field is plotted to show the Dirac cones, where the value of $S_z$ is maximal. This reveals that the intralattice hopping generates a gap in the energy spectrum.}
  \label{FigTextures}
\end{figure}

The bipartite nature of the honeycomb lattice allows us to separately probe the atomic population densities for each hyperfine state. We can compare these values at each point of the Brillouin zone to obtain a field $S_z(\vc{k})=a^\dagger_ka_k-b^\dagger_kb_k$. Moreover, an adiabatic protocol such as the one described in the previous subsection or in-flight Raman-assisted internal state rotations provide a way of measuring the ``rotated'' fields $S_x(\vec{k})=a^\dagger_kb_k+b^\dagger_ka_k$ and $S_y(\vec{k})=i(a^\dagger_kb_k-b^\dagger_ka_k)$. This observable vector field on the Brillouin zone ($\vc{S}(\vc{k})=(S_x,S_y,S_z)$) is a powerful tool to analyze some distinct features of the ground state: the presence of a gap in the energy band structure ($S_z(\vc{K}_\pm) \neq 0$), the characteristic state differences between cones or even the topological nature of the ground state~\cite{Alba11}. In figure~\ref{FigTextures} we show a simulation of such a measurement in two distinct cases: figures~\ref{FigTextures}a,~\ref{FigTextures}b,~\ref{FigTextures}c feature the value of the phase $\phi=\tan^{-1} (S_y/S_x)$ for different values of the Abelian potential $\vc{A}$. The displacement of the cones provoked by $\vc{A}$ along the Brillouin zone is apparent, as it is the fact that the cones are vortices in the $S_x,S_y$ vector field. Figure~\ref{FigTextures}d illustrates how a gap or effective mass created by a non-zero value of $\Gamma_A$ can be observed by measuring $S_z$. Once again, the Brillouin zone and the $S_x,S_y$ vector field are depicted to illustrate the vortex effect of the Dirac cone.

\subsection{Few particle dynamics}
\label{sec:klein}

The method in the previous subsection gives us access, among other things, to the best known experimental observable in condensed matter physics, which is the density of states. It has been the easiest to measure and therefore has become the default choice in optical lattice simulations of solid state physics. However, the distinctive characteristics of the excitations in a half-filled hexagonal lattice may be worth an extra effort: trying to observe the behaviour of a group of particles with well-defined momentum obeying a complete Dirac equation will reveal some of the distinctive features of the simulated fields.

As an example we suggest using the Klein tunneling effect~\cite{Klein29} to probe and measure the energy gaps between the two bands. Let us add a uniform electric field pointing along one direction, say $x$, described by a linearly growing potential, $V x$. As explained in Ref.~\cite{Casanova10}, the effect of this potential is to accelerate particles, continuously increasing their momenta in time,
\begin{equation}
\label{rampspeeds}
k_x(t) = k_x(0) - Vt/\hbar c,
\end{equation}
until the particle reaches a boundary of the Brillouin zone. We have set an adimensional time scale $t=(time) \times J/\hbar $. At this point two things may happen. If the particle is far away from a Dirac singularity, or the minimum gap between energy bands (the effective mass $m$) is large compared to the acceleration, $mc^2 \gg V,$ the particle will simply reappear through the opposite side of the Brillouin zone, reversing its velocity and performing the so called Bloch oscillations. However, if the particle hits against the proximities of the $K_\pm$ points and the gap is small, $m c^2 \ll V,$ the particle will experience a Landau-Zener process and jump to the opposite energy band, maintaining its group velocity.

\begin{figure}
  \centering
  \includegraphics[width=\textwidth]{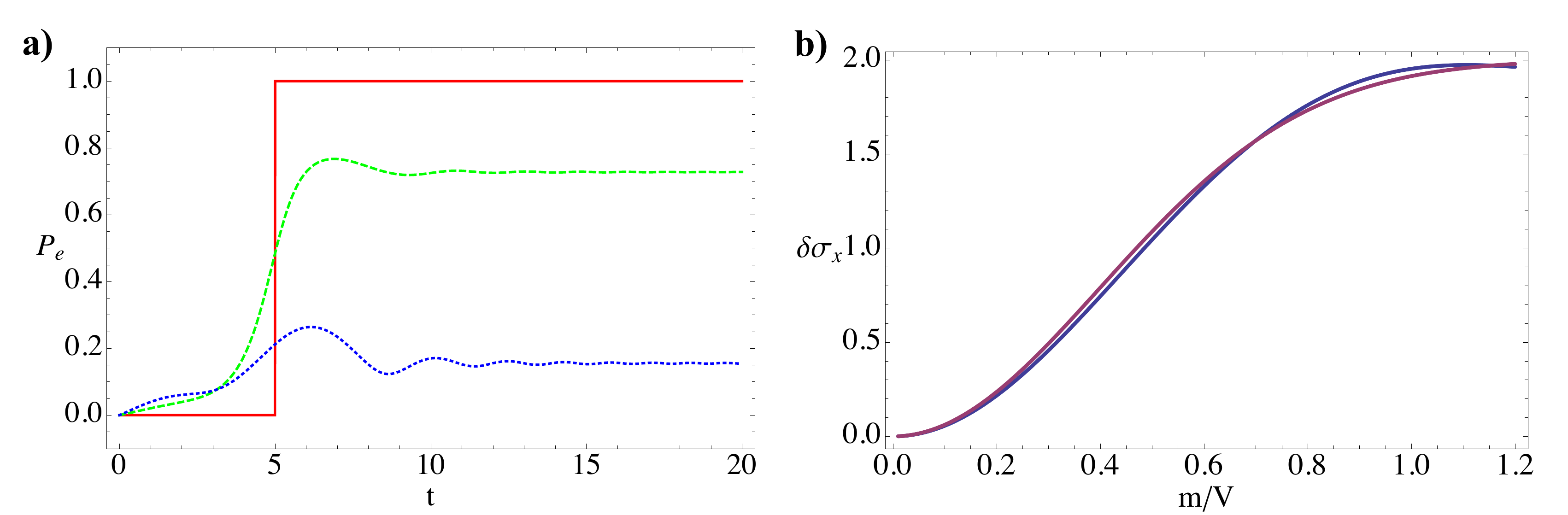}
  \caption{a) Excitation probability for three different ramp speeds V (equation~(\ref{rampspeeds})): No gap
    ($m$=0, solid line),
    medium gap ($m$=$V$, dashed line) and large gap ($m=2.5V$, dotted line) against time. While the gapless
    state gets immediately promoted once it reaches the Dirac points,
    the gapped states have reduced positive energy contributions. b) Change in $\langle \sigma_x \rangle$: $\langle \sigma_x
    (t=\infty) \rangle - \langle \sigma_x (t=-\infty) \rangle$ depending
    on the value of the effective mass. We compare the change in
    $\langle \sigma_x \rangle$ by an exact simulation of
    the particle dynamics with the Landau-Zener formula (purple). We denote by $t=+(-) \infty$ the times where the particle is still (already) away from the cone. }
  \label{ProbExc}
\end{figure}

As shown in figure~\ref{ProbExc}, we have simulated numerically this process for different masses of the Dirac field. As a signature of the jump between bands we simply use the expectation value of $\vc{k} \cdot \SIGMA,$ which is directly correlated to band excitation probability but easier to measure. Note how for the gapless phase, $m=0$, there is a perfect jump (Figure~\ref{ProbExc}a), and how the probability is well approximated by the Landau-Zener formula (Figure~\ref{ProbExc}b), which allows us to reverse-engineer the experiment and fit the value of $m$. 

How would this be implemented in an experiment? The idea would be to do the same optical lattice setup with a small number of fermions cooled down to the lowest value of the momenta, $\vc{k}=0$, in a honeycomb lattice that implements the desired Dirac Hamiltonian, with or without effective mass. One would then activate an electric field along the direction $\vec w$, for a certain time $t$. After this time one would measure the state of the atoms, or more precisely the expectation value $\langle \SIGMA\rangle.$ By changing the duration of the field and its intensity, and monitoring the changes in $\vc{k} \cdot \SIGMA $, one would be able to reconstruct not only the Klein effect but also the whole spin texture of the bands, as discussed in the previous subsection.

\section{Conclusions}
\label{sec:discussion}

In this work we have presented an experimental proposal to simulate Dirac fermions interacting with an effective gauge field. The fermionic component of the model is obtained by trapping atoms in two triangular optical lattices that are connected to form a honeycomb lattice~\cite{Semenoff84}. As explained in the manuscript, the low-energy excitations of such a model may be described using an effective theory that consists of two flavors of non-interacting Dirac fermions. The gauge fields, on the other hand, arise from perturbations of the atom dynamics, such as lattice distortions, short- and long-wavelength modulations of the hopping amplitudes and state-dependent external potentials. Such perturbations are particularly easy to implement using our setup, which consists of two independent lattices.

Our work connects both with recent developments in the field of graphene~\cite{Castro08,Geli10,Castro10} and with the field of quantum simulation of synthetic gauge fields~\cite{Jaksch03,Spielman09,Mazza11,Jiannis09}, but with various advantages. On the implementation side, the optical lattice setup allows for a single-site resolution and a local customization of the potentials which is hard to think of in solid state implementations. Moreover, the use of two atomic species in a bipartite lattice introduces new measurement possibilities, such as the direct observation of the fermionic fields [See section~\ref{sec:textures}], or the study of state-dependent Bloch oscillations~[Section~\ref{sec:klein}]. In comparison with other ultracold atom proposals, while we still rely on the use of assisted tunneling, ours is a static and straightforward setup, where tunneling is implemented by a simple optical field without hopping unitaries, and which nevertheless allows for realistic values of the couplings.

The proposal and the study in this work are also of theoretical interest. The work with optical lattices allows us to make an accurate and rigorous connection between the microscopic theory of trapped atoms and the simulated quantum field theory. While a similar work has been done for strained and curved graphene sheets~\cite{Geli10}, the optical lattice setup allows us to compute from first principles how the microscopic changes in the optical potential lead to ``strain'' and hopping distortions. In some cases, as in section~\ref{sec:raman-distortion} a simple modulation of the Raman lasers translates into a similar modulation of the hoppings, giving rise to the gauge fields. In other cases, we may concentrate on geometric deformations of the trapping potential and  rigorously work out how they affect the hopping matrices, in a theoretically pleasant and flexible way. We foresee that the same tools developed in this manuscript will also help solve an open problem in the graphene world, which is the relation between microscopic deformations of the honeycomb lattice and the appearance of an effective metric and curvature. Instead of introducing an effective spin connection by hand, as it is currently done for carbon layers~\cite{Geli10}, we expect to be able to derive the same metric directly from deformations of the optical potential.

Finally, it is also remarkable the fact that our models allows the introduction of on-site and nearest-neighbor atomic interactions at no cost and with realistic values~[Section~\ref{sec:wannier}]. At this point the model stops being a single-particle theory and becomes numerically intractable, entering the regime in which quantum simulation provides both new problems and interesting answers.

\section*{Acknowledgements}

This work has been funded by Spanish MICINN Project FIS2009-10061, FPU grant No.AP 2009-1761, CAM research consortium QUITEMAD S2009-ESP-1594, a Marie Curie Intra European Fellowship, CSIC Grant JAE-INT-1072, the POLATOM network and the Royal Society.

\section*{References}

\bibliographystyle{unsrt}

\end{document}